# SMT-Based Bounded Model Checking of Fixed-Point Digital Controllers

Iury Bessa, Renato Abreu, João Edgar Filho, and Lucas Cordeiro
Electronic and Information Research Center, Federal University of Amazonas, Brazil
E-mails: {iurybessa,renatoabreu,lucascordeiro,jo_edgar}@ufam.edu.br

*Abstract*— Digital controllers have several advantages with respect to their flexibility and design's simplicity. However, they are subject to problems that are not faced by analog controllers. In particular, these problems are related to the finite word-length implementation that might lead to overflows, limit cycles, and time constraints in fixed-point processors. This paper proposes a new method to detect design's errors in digital controllers using a state-of-the art bounded model checker based on satisfiability modulo theories. The experiments with digital controllers for a ball and beam plant demonstrate that the proposed method can be very effective in finding errors in digital controllers than other existing approaches based on traditional simulations tools.

*Keywords*— *Digital controllers, fixed-point, direct forms, model-checking*

## I. Introduction

Nowadays, almost all control systems are implemented in computational structures increasing the applications of digital controllers. Digital controllers have improved the flexibility of control algorithms, since a controller may be implemented with different software variations using the same hardware structure; this reduces the design time and consequently simplifies the design process. Digital controller designers do not exploit all the computer implementation advantages if they only reproduce the traditional analog techniques (e.g., PID and lag/lead control) in a computer-based system [1]. To achieve the best advantages of computational implementation, the computer-controlled system must exploit all digital control techniques. However, this might lead to problems related to finite word-length realizations, which represent an important area of research in the control system community.

Digital controllers are typically implemented in microcomputers, microprocessors, or digital signal processors. Any digital computer with a data acquisition system and an operating system can be used to implement a digital controller. These implementations might use fixed-point or floating-point arithmetic. Since floating-point implementation has a greater number of representable values and consequently reduced errors, the fixed-point processors are the fastest and cheapest and consequently, they are more common in practice. In this context, problems caused by finite word-length have greater dimensions (i.e., quantization and overflow errors); and control systems are thus subjects to problems that only occur in digital controller realizations. These problems could be fixed or at least reduced according to the chosen computational structure (e.g., direct forms), which could increase or decrease the number of arithmetical operations and quantizations effects.

Additionally, there is another major problem that might occur in digital controller realizations, which is related to time constraints. Digital controllers are strictly real-time systems. The controls tasks execution cannot take more time than a sample period chosen by the control engineer. Hence, the controller's implementation must consider the code execution time and the sample time compatibility. Control engineers are, in principle, aware about these problems; but they frequently use simulation tools to validate their controllers and to check whether the desired performance is achieved. However, most simulation tools, e.g., PSIM [2], LABVIEW [3], and MATLAB [4], are based on floating-point arithmetic and thus ignore all problems that might occur in fixed-point implementations. There are some tools that simulate fixed-point systems, but they show poor results, because they neither cover all possible scenarios nor check time constraints, which are important in real-time systems [5].

An example of simulation tool is proposed by Sung and Kum, where an algorithm is developed to determine the minimum bound of the word-length fixed-point representation through simulation methods [5]. However, as any other simulation tool, it cannot explore all possible scenarios and thus problems might not be detected. An interesting work is presented by Anta et al. [6], where a tool called Costan is developed. Costan finds errors in implementation of a mathematical model and verifies whether the error is tolerated, considering the quantization effect and fixed-point implementation; and then focuses its analysis in the stability of the system. In particular, Costan verifies the C implementation of the controller and checks the maximum possible error between the C model and the SIMULINK model of the controller via a symbolic error analysis. Some recent work uses a formal verification methodology based on bounded model checking (BMC) with satisfiability modulo theory (SMT) solvers. Cox et al. show that simulations tools are useful, but insufficient [7]; the authors propose the use of SMT-based BMC to verify digital filters. Most recently, Abreu et al. verify various digital filters properties (e.g., overflows, limit cycles, times constrains, stability, and frequency response) using a state-of-the-art BMC tool, called ESBMC (Efficient SMT-based Bounded Model Checking) [8].

In this paper, digital controllers' implementations are verified by using an SMT-based BMC tool; similar to Cox et al. and Abreu et al. [7], [8]. In fact, a digital controller can be seen as a form of filter, but in digital controllers, all actions must happen in real-time [9] and this differentiates this work from others. Underflow and overflow that can occur in additions or multiplications operations are considered using different realization structures of digital controllers. Additionally, the occurrence of limit cycles and the stability of digital controllers are verified in a commercial industrial plant.

## II. Background

This section describes some digital control systems concepts, structures as well as implementation problems caused by the use of a fixed-point processor. The satisfiability modulo theory and BMC concepts are also addressed here.

## A. Fixed-Point Digital Controllers Implementation

A digital controller is a linear time-invariant causal discrete-time dynamic system [9]. A digital controller manipulates discrete numerical signals; and its implementation is a program executed by a microprocessor. There are several ways to represent a digital controller. A popular manner is the mathematical representation through transfer function in the $Z$ domain [10] as follows

$$G(z) = \frac{b_0 + b_1 z^{-1} + \cdots + b_M z^{-M}}{1 + a_1 z^{-1} + \cdots + a_N z^{-N}}. \quad (1)$$

Another mathematical representation of digital controller is the difference equation, which may be described as

$$y(n) = -\sum_{k=1}^{N} a_k y(n-k) + \sum_{k=0}^{M} b_k x(n-k), \quad (2)$$

where $y(n)$ is the output in instant $n$ and $x(n)$ is the input in instant $n$ [10]. There are many techniques for digital controllers' design. These design techniques are a huge topic that is not covered here, but there are several references that provide this background, e.g., [1], [9], [11]-[13].

There are many ways to implement a digital controller in software; and how the controller organization is implemented, will influence its performance. Different realizations of digital controllers are studied in several books [11]- [14]. In this work, however, only direct forms implementations are considered. Direct realizations use the coefficients of Equation (2) and this is the major disadvantage, since it makes the implementation extremely sensitive to numerical errors, especially in the finite word-length implementation with fixed-point. The advantage of this implementation is that states variables are derivations of delayed inputs and outputs [15]. Three direct structures implementations are exploited: Direct Form I (DFI), Direct Form II (DFII), and Transposed Direct Form II (TDFII). Fig.1 shows the DFI; other structures may be studied with details in [15] and [16]. In general, the implementation of digital controllers is subject to various limitations (e.g., e.g., quantization and operations round-off) that must be understood and their consequences estimated, especially when the implementation occurs in a fixed-point processor [15].

In particular, a quantizer approximates a signal value by a value from a discrete finite set, generating a rounding error, whose maximum value will be assumed by $2^{-l-1}$, where $l$ is the number of bits of the fractional part. The quantization in the finite word-length operations often causes periodic oscillations known as limit cycles, which are caused by round-off errors in multiplication and overflow errors in addition [15].

The overflow occurs when a sum or product is outside the range of representable numbers. There are two main ways of overflow handling: wrap-around and saturation. The first way ignores the overflow, allowing the numerical representation of a result to be greater than a maximum representable value to be stored with the least significant bits only (i.e., it wraps). The second way, hold the maximum representation value when overflow occurs [17]. All these problems are known as Finite Word Length (FWL) effects. A realistic model of a FWL system must include the quantization of every numerical value, including arithmetic results, input signals, and system coefficients. An example of realistic representation of a third-order system implemented in TDFII is shown in Fig. 2.

The typical fixed-point representation uses two-complement to represent signed binary values. A standard representation of a fixed point number is $<k, l>$, where $k$ represents the number of bits of the integer part, and $l$ represents the number of bits of the fractional part. The most significant bit is the sign bit; therefore, the representable range of values is between $2^{k-1} - 2^{-l}$ and $-2^{k-1}$.

Naturally, the FWL effects are more present in fixed-point implementations. There are several approaches that aim to minimize these effects in fixed-points processors. However, traditional tools for simulation and testing do not appear to be sufficient in validation of fixed-point digital controllers' implementation, because they explore a limited number of scenarios and values. From the literature review, it is evident that there is some need for a formal verification method for digital controllers.

## B. SMT-Based Bounded Model Checking

The basic idea of BMC is to check (the negation of) a given property at a given depth. Supposing a transition system $M$, a property $\phi$ and a bound $k$, BMC unrolls the system $k$ times and translates it into a verification condition (VC) $\psi$, in such a way that $\psi$ is satisfiable if and only if $\phi$ has a counterexample, of depth less than or equal to $k$; thus, standard SMT solvers can be used to check whether $\psi$ is satisfiable.

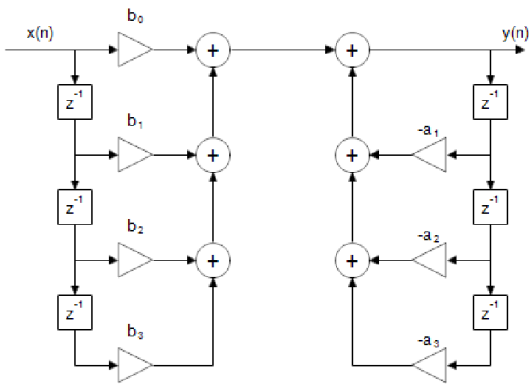

Fig. 1 Direct Form I Realization

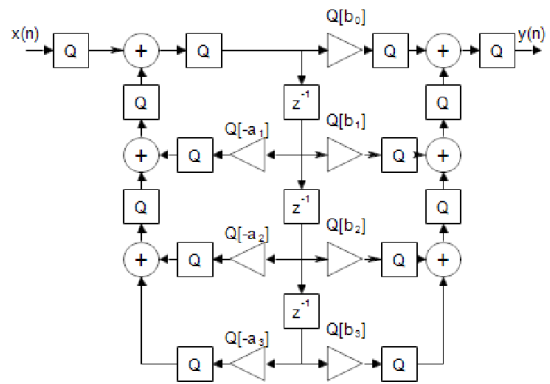

Fig. 2 Realistic model of a third-order digital controller

In BMC of digital controllers, the bound $k$ limits the number of loop iterations and recursive calls in the controller implementation. BMC thus generates VCs that reflect the exact path in which a statement is executed, the context in which a given function is called, and the bit-accurate representation of expressions [18]. In this work, the ESBMC tool is used as a verification engine, since it represents one of the most efficient BMC tools in the last software verification competitions [12], [13]. In ESBMC, the associated SMT-based BMC problem is formulated by constructing the following logical formula

$$\psi_k = I(s_0) \wedge \vee_{i=0}^{k} \wedge_{j=0}^{i-1} \gamma(s_j, s_{j+1}) \wedge \overline{\phi(s_i)}, \quad (3)$$

where $\phi$ is a safety property (e.g., overflow), I is the set of initial states of $M$, and $\gamma(s_j, s_{j+1})$ is the transition relation of $M$ between time steps $j$ and $j+1$. Hence, $I(s_0) \wedge \wedge_{j=0}^{i-1} \gamma(s_j, s_{j+1})$ represents the executions of a transition system $M$ of length $i$. The above VC $\psi_k$ can be satisfied if and only if, for some $i \leq k$, there exists a reachable state, at time step $i$, in which $\phi$ is violated. If Equation (3) is satisfiable, then the SMT solver provides a satisfying assignment, from which the values of the controller variables can be extracted, in order to construct a counterexample. The latter, for a property $\phi$, is then defined as a sequence of states $s_0, s_1, \ldots, s_k$ with $s_0 \in S_0$, $s_k \in S$ and $\gamma(s_i, s_{i+1})$, for $0 \leq i < k$; and this can be used to reproduce the error in traditional simulation-based tools. If Equation (3) is unsatisfiable, then one can concluded that there is no error state in $k$ steps or less.

### III. VERIFICATION OF DIGITAL CONTROLLERS

To explain the verification of digital controllers, the ball and beam discrete model is used as a running example [19]- [20]. The digital controllers for a Quanser's ball and beam plant with SRV02 actuator set are properly designed; all plant parameters and mathematical models are extracted from user manuals.

As a first step, controllers can be designed through different techniques, e.g., emulation, Ragazzini, Truxal, and discretization [11], [12], and [9]. Secondly, after designing the controllers, their behaviors can then be simulated in SIMULINK, which is part of the MATLAB toolset [4]. Here, the closed-loop responses are verified from simulations to check the step-response of the system; when necessary, other types of testing signals (e.g., ramp or parable) are also applied to the control system. Thirdly, after the simulation, the output range for a specific input is estimated, and the word-length of the fixed-point representation is chosen. Fourthly, once the word-length and the transfer function of the controller are obtained, the digital controller is then implemented in a C model for a specific fixed-point microprocessor architecture with a known clock time; it allows the analysis of the digital controller behavior in the time domain. The final step of the proposed method is the verification of the properties. For that, assertions are inserted in the C model of the controller to check for four particular properties: overflow, limit cycle, stability, and time constraint. The verification of these properties are carried out by ESBMC, which checks the implementation of the controller according to its specification, even if the properties (extracted from the specification) do not require an exhaustive checking via non-deterministic inputs [8].

The ESBMC aids the control engineer to optimize their controllers' implementation; in particular, it helps them choose the sample time, quantization range, word-length, and implementation structure. When a property violation is detected, the control engineer acts by fixing the identified problem in the controller's design. For instance, when an underflow or overflow occurs, an output error violation will occur too, and the engineer must perform a new verification with the same controller (and the same poles and zeros positions), but with a reduced gain or with a larger word-length. However, if a time constrain violation is detected, the engineer must reduce the word-length, and if the problem persists, the controller has to be redesigned with a lower complexity or with a greater sample time, if it does not affect the system stability. Model checking digital controllers is thus an interactive process, whereby the engineer should fit the controller mathematical representation to the given microprocessor architecture, finding the optimal fixed-point representation, and thus avoiding implementation problems, which are typically met in the physical implementation and whose causes are hard to be detected.

### A. Arithmetic Underflow and Overflow Verification

The arithmetic underflow and overflow verification without a computational tool is a very challenging task; BMC tools appear to be a good solution for this. In this work, the quantizer C code contains assertions and ESBMC is thus configured to detect underflow and overflow in a digital controller with a specific fixed-point word-length through the application of non-deterministic inputs to the already quantized mathematical model. For any addition or multiplication results, during controller operation, if there exists a value that exceeds the range representable by the fixed-point, a VC detects it as an underflow or overflow violation. Here, a literal $l_{overflow}$ is generated in order to represent the validity of each addition and multiplication operation, according to the following constraint

$$l_{overflow} \Leftrightarrow (MIN \leq FP) \wedge (FP \leq MAX), \quad (4)$$

where *FP* is the fixed-point representation for the result of the adders and multipliers after the quantization, and *MIN* and *MAX* are, respectively, the minimum and maximum values of the representable range for the given fixed-point bit format. A failed overflow verification example is shown in Table 1. Here, a controller (see test case 9 of Table 3) is verified with the DFI realization. The fixed-point representation format is $<4,11>$ and the input range is [-6,6]. However, the sequence of inputs in Table 1 leads the output to a number that is greater than the representable limit, thus occurring the overflow. The verification engine indicates that failure, and gives as counterexample the sequence of inputs shown in Table 1, which can be easily reproduced using the difference equation to compute outputs values; note that this particular defect may be unnoticed by simulation tools (e.g., Matlab) unless one knows the input sequence that leads to the overflow, which is not generally the case.

Table 1. Overflow failure example

| n | 1 | 2 | 3 | 4 | 5 |
|---|---|---|---|---|---|
| $x(n)$ | 6.0000 | 5.9990 | −5.9990 | 6.0000 | 5.9995 |
| $y(n)$ | 0.6 | −1.6801 | 2.5025 | −4.3369 | 12.1032 |

## B. Limit Cycle Verification

The steady state response of a control system is the portion of total response that remains after the transient effect becomes insignificant [21]. In this way, the step response of a stable control system should be a constant value after a certain time. However, when the limit cycle occurs, it is not necessarily true. The limit cycle phenomenon consists in the presence of oscillations occurring in the output, even when the input sequence is a constant value [15]. These oscillations may be very harming to the control systems, because they may cause damages to the physical system (especially in mechanical systems) and harm surround products [22].

To verify the limit cycle occurrence in a digital controller, the quantization process wraps around when the overflow occurs. Thus, the verification engine does not detect overflow failures. For the limit cycles test, the verification engine is configured to input a zero sequence and initialize the system with a non-deterministic initial state. A verification condition is then added to detect the limit cycle failure, i.e., it detects a failure if a sequence of outputs states are repeated during the zero inputs sequence.

An example of failure in limit cycle verification is shown in Fig. 3. This is a digital controller (see the test case 11 in Table 3) in DFI realization, with output range of $[-4,4]$ and with fixed-point representation $<2,13>$. The verification engine checks the failure occurrence and gives the following counterexample: if the system receives a zero sequence, following a $\{2,2,2,2\}$ sequence of past outputs, the limit cycle will occur, as shown in the graph of Fig. 3. In this graph, a simulation with 2 seconds of duration is shown, reproducing the counterexample provided by the verification engine.

## C. Time Constrains Verification

The sample time is a very important parameter to be chosen in a digital control system. In particular, all the system's dynamic is changed with a modification in the sample time. A precise selection of sample period is thus essential for a computer-controlled system. On the one hand, too short sample times require a greater processing performance and consequently processors with a high clock frequency; this can impose technical limitation in the design of the digital controller. On the other hand, too long sample times do not permit the reconstruction of continuous signals [11]. In principle, the sample time choice depends on the plant, where the control system is applied. The right choice of the computational implementation of a controller may thus reduce the number of arithmetic operations and consequently the computational costs. As control systems are typically real-time systems, they cannot take more time to process tasks than a sample period. In practical applications, the controller is designed with a reasonable sample period, which shows good simulations results. Thereafter, it is implemented in a computer system, where samples are scheduled at every sample period; this is the maximum time that the processor takes to perform all control tasks and operations. If an operation cannot terminate on time, then the results might not be correct and the control system might not work as expected.

For this particular reason, a time constraint verification tool becomes a very useful controller design tool, which may indicate if the chosen sample period and the computational realization are compatible, before the physical implementation, thus avoiding serious malfunctions of the system.

As a result, the needed time to execute a specific code can be estimated, once each instruction can be broken into a set of assembly instructions; in particular, every processor has a table of clock cycles spent on each assembly instruction. To know the total time needed to execute a code, the number of clock cycles must be divided by the processor clock rate (or multiplied by the clock time). However, the estimation of clock cycles is a challenging task, once a controller's implementation contains loops and decision statements, which can take different number of clock cycles to execute, depending on the input parameters (that are usually non-deterministic values). In order to verify time constraints, a literal $l_{timing}$ is generated to represent the time response, with the following constraint

$$l_{timing} \Leftrightarrow ((N \times T) \leq D), \quad (5)$$

where $N$ is the number of cycles spent by the digital controller, $T$ is the clock period and $D$ is the deadline [8].

## D. Poles and Zeros Verification

The stability of a system may be verified through the positioning of its poles. A discrete system is stable if all its poles are in the interior region of the unitary circle of $z$-plane, i.e., the poles must have the module less than one [16]. Thus, the stability verification of a system should be done with an algorithm that determines the roots of the transfer function denominator polynomial.

In this work, the Eigen Library [23] is used, in order to determine the roots of a polynomial. The three steps of the algorithm can be described as follows:

1. Given a polynomial $p(t) = t^n + a_{(n-1)}t^{n-1} + \cdots + a_1 t + a_0$, determine the companion matrix $A$, such that:

$$A = \begin{bmatrix} 0 & 1 & 0 & \cdots & 0 \\ 0 & 0 & 1 & \cdots & 0 \\ \cdots & \cdots & \vdots & \ddots & \vdots \\ 0 & 0 & 0 & \cdots & 1 \\ -a_0 & -a_1 & -a_2 & \cdots & -a_{n-1} \end{bmatrix};$$

2. Reduce the matrix $A$ to the Real Schur form;
3. Apply the Schur Decomposition to compute the roots of polynomial $p(t)$.

ESBMC checks the system's stability, by verifying whether all eigenvalues show absolute values less than one, after the coefficients quantization. If any eigenvalue absolute value is greater than one, then stability fails and a counterexample is reported.

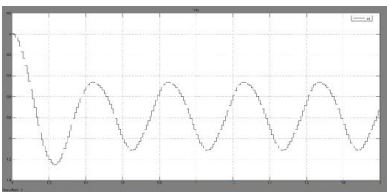

Fig. 3. Limit Cycle in a Digital Controller

Table 2. Digital Controllers for a Ball and Beam Plant

| # | Numerator | Denominator | $t_s$ | OVS | $e_{ss}$ |
|---|---|---|---|---|---|
| A | [0.15 0.05 0.40] | [1.0 0.0 0.3] | Inf. | Inf. | Inf. |
| B | $[2.0\ -4.0\ 2.0] \times 10^5$ | $[1.00\ 0.00\ -0.25]$ | 0.35 | 27% | 0 |
| C | $[50.000\ -140.950\ 131{,}850\ -40.935]$ | $[1.00000\ -1.97000\ 1.03300\ -0.06068]$ | 7.24 | 0 | 0 |
| D | $[9.37\ -35.82\ 52.01\ -3.482\ 10.03\ -0.78] \times 10^8$ | $[1.000\ 9.112\ -2.247\ -8.656\ 0.657\ 0.135]$ | Inf. | Inf. | Inf. |
| E | $[1.0\ -3.0\ 3.0\ -1.0] \times 10^7$ | $[1.000\ 1.800\ 1.140\ -0.272]$ | 0.12 | 57% | 36% |
| F | $[1.0\ -2.5\ 2.0\ -0.5] \times 10^5$ | $[1.000\ 1.500\ 0.680\ 0.096]$ | 2.62 | 0 | 0 |

## IV. EXPERIMENTAL EVALUATION

This section is split into three parts. The first section describes all digital controllers that are designed for the Ball and Beam plant. The second section describes the experiments configuration and the last section summarizes the results.

### A. Digital Controllers' Design for a Ball and Beam Plant

Digital controllers for a Quanser ball and beam plant are developed using different techniques with MATLAB's aid, as described in Section II; and they are all simulated in SIMULINK. The objective of this control system is to stabilize the ball in a desired position along the beam; for that purpose, the controller should input a voltage signal in the SRV02's system, which rotates the beam by adjusting its angle. From the specification, the plant parameters and model are extracted (Quanser, 2008). The discrete form of the plant, using a sample time of $0.01\ s$, is given by

$$G(z) = \frac{1.0067 \times 10^{-8}(z+9.256)(z+0.9324)(z+0.9389)}{(z-1)^3(z-0.7041)}.$$

Controllers with different performances are designed and simulated in SIMULINK. Table 2 describes the controllers with these numerators and denominators vectors, and a summary with simulation results, i.e., settling time ($t_s$), overshooting ($OVS$), and steady-state error ($e_{ss}$).

### B. Experimental Setup

For the following verifications, a 16-bits microcontroller with a clock rate of 16 MHz is used as the embedded platform, where the controllers are actually implemented; all sample rates are adjusted to 100 Hz. Table 3 summaries different controllers' configurations. A physical implementation with a signal conditioning circuit external to the microcontroller with an external gain is assumed. To understand the influence of the realization structures on overflow and underflow, limit cycle, and time constrains, all controllers are implemented in three different realizations: DFI, DFII, and TDFII. These realizations have no effect in the system's stability, once it only checks the effect of coefficient quantization and round-offs on poles and zeroes of the digital controllers. Note that the second-order structures (i.e., parallel or cascade) are not addressed here, but only direct implementations that are more susceptible to errors.

This work employs ESBMC v1.23[1], with the SMT solver Z3 v4.0. All tests are executed with a maximum verification time of 3600s. If the time needed to finish the verification is greater than this maximum, then the verification is aborted. ESBMC is invoked by setting the file name, timeout, and the SMT solver. Additionally, division by zero, array bounds, and pointer safety verifications are disabled, once the main objective is the controller properties checking. The experiments are executed in a computer with the following hardware configurations: Intel Core i7-2600 3.40 GHz processor, 24 GB of RAM, Ubuntu 11.10 Maverick Meerkat 64-bits OS.

### C. Experimental Results

Table 3 presents the verification results. Here, *S* represents a successful test and *F* represents a failed test. If the verification exceeds the limit time, then the result is represented by TO. According to the experimental results, ESBMC detects various errors in different realizations of digital controllers. However, the verification process takes a time that may be longer, if the controller order is higher. Others factors that may influence this time is the precision of fixed-point implementation; if the number of fractional bits is increased, then the verification time tends to increase as well. Furthermore, in the limit cycle tests, the length of zero input vectors used to verify oscillations occurrence must be greater or equal than the length of the fractional part, i.e., the limit cycle verification time will be much longer if the precision is greater. Typically, the successful verifications tend to take more time than failed ones, once the verification process only stops when an error is found or when all VCs are satisfied.

The results also points out that ESBMC is a useful tool to determine the better fixed-point structure realization for digital controllers; for example, analyzing the results in Table 3 (lines 1, 2, 3 and 10), a control engineer may easily conclude that the controllers A and B should be implemented in the DFI or DFII instead of the TDFII, in order to avoid limit cycle oscillations. Furthermore, some failures that appear in the counterexamples are difficult to be found by simulation tools. As an example, one can analyze the stability of a closed-loop control system using the controller C in SIMULINK; and presume that the closed-loop system will be stable. This controller is designed by emulation and mapping of analogs poles and zeroes with the following zero-poles-gain representation:

$$C_3(z) = \frac{50\big((z-1)^2(z-0.81873)\big)}{(z-0.9704)(z-0.9329)(z-0.06\ 7032)}.$$

Two zeroes on 1 can be observed in this controller to cancel two poles in 1 of the ball and beam plant, and then stabilize the closed-loop system. When this closed-loop system is simulated, the poles and zeroes cancellation occurs and the system's response is acceptable (the step response is shown in Figure 4). However, if the transfer function with quantized coefficients is simulated, then the response is totally different (see Figure 5). When the closed-loop system model is verified by ESBMC, the stability test fails due to the non-cancellation of unstable poles on 1; the cancellation does not occur due to errors caused by the FWL effects. Some other examples of reduction of controller's precision are described by Satina et al. [24].

---

[1] The ESBMC tool and benchmarks are available at *www.esbmc.org*

Table 3. Experimental Results

| # | Controller | Gain | Input Range | Bits | Type | Overflow Result | Overflow Time | Limit Cycle Result | Limit Cycle Time | Timing Result | Timing Time | Stability Result |
|---|---|---|---|---|---|---|---|---|---|---|---|---|
| 1 | A | 1 | [−1,1] | <3,4> | DFI | S | 19.8 | S | 32.9 | S | 0.6 | S |
|   |   |   |   |   | DFII | S | 15.7 | S | 235.8 | S | 0.6 |   |
|   |   |   |   |   | TDFII | S | 79.0 | F | 102.2 | S | 0.6 |   |
| 2 | B | $10^6$ | [−1,1] | <2,6> | DFI | F | 1.7 | S | 62.2 | S | 0.6 | S |
|   |   |   |   |   | DFII | F | 1.6 | S | 252.0 | S | 0.6 |   |
|   |   |   |   |   | TDFII | F | 1.6 | F | 114.8 | S | 0.6 |   |
| 3 | B | $10^7$ | [−1,1] | <4,3> | DFI | S | 22.0 | S | 23.0 | S | 0.6 | S |
|   |   |   |   |   | DFII | S | 10.2 | S | 131.9 | S | 0.6 |   |
|   |   |   |   |   | TDFII | S | 59.1 | F | 179.6 | S | 0.6 |   |
| 4 | C | 50 | [−1,1] | <2,13> | DFI | F | 79.2 | TO | - | S | 0.6 | S |
|   |   |   |   |   | DFII | F | 29.7 | F | 686.8 | S | 0.6 |   |
|   |   |   |   |   | TDFII | F | 131.4 | TO | - | S | 0.6 |   |
| 5 | D | $10^9$ | [−1,1] | <2,13> | DFI | F | 1771,7 | TO | - | S | 0.7 | F |
|   |   |   |   |   | DFII | F | 437.5 | TO | - | S | 0.7 |   |
|   |   |   |   |   | TDFII | F | 2085.2 | TO | - | S | 0.7 |   |
| 6 | D | $10^{10}$ | [−1,1] | <2,13> | DFI | F | 3437.2 | S | 14.8 | S | 0.7 | F |
|   |   |   |   |   | DFII | F | 860.0 | S | 28.9 | S | 0.7 |   |
|   |   |   |   |   | TDFII | F | 2522.7 | S | 25.8 | S | 0.7 |   |
| 7 | C | 500 | [−4,4] | <2,13> | DFI | F | 102.0 | S | 5.6 | S | 0.6 | S |
|   |   |   |   |   | DFII | F | 34.5 | S | 20.0 | S | 0.6 |   |
|   |   |   |   |   | TDFII | F | 555.5 | S | 9.4 | S | 0.6 |   |
| 8 | C | 500 | [−5,5] | <2,8> | DFI | F | 48.6 | F | 494.3 | S | 0.6 | S |
|   |   |   |   |   | DFII | F | 24.3 | TO | - | S | 0.6 |   |
|   |   |   |   |   | TDFII | F | 190.5 | TO | - | S | 0.6 |   |
| 9 | C | 500 | [−6,6] | <4,11> | DFI | TO | - | TO | - | S | 0.6 | S |
|   |   |   |   |   | DFII | F | 12.8 | TO | - | S | 0.6 |   |
|   |   |   |   |   | TDFII | TO | - | F | 2503.6 | S | 0.6 |   |
| 10 | B | $10^7$ | [−1,1] | <3,12> | DFI | S | 25.1 | S | 334.2 | S | 0.6 | S |
|   |   |   |   |   | DFII | S | 19.6 | S | 1122.6 | S | 0.6 |   |
|   |   |   |   |   | TDFII | S | 68.7 | F | 250.1 | S | 0.6 |   |
| 11 | E | $10^7$ | [−4,4] | <2,13> | DFI | F | 352.4 | S | 5.9 | S | 0.6 | S |
|   |   |   |   |   | DFII | F | 55.7 | S | 13.3 | S | 0.6 |   |
|   |   |   |   |   | TDFII | F | 178.0 | S | 10.0 | S | 0.6 |   |
| 12 | F | $10^7$ | [−2,2] | <2,13> | DFI | F | 14.9 | S | 5.6 | S | 0.6 | S |
|   |   |   |   |   | DFII | F | 11.3 | S | 11.9 | S | 0.6 |   |
|   |   |   |   |   | TDFII | F | 77.5 | S | 8.8 | S | 0.6 |   |

Note that the stability verification time is not shown in Table 3, since they are very fast to be checked, (i.e., each verification run takes less than one second). The results show that limit cycles failures occurs more frequently in DTFII structure than others studied here; however, this structure presents less arithmetical operations, which means less computational effort and less chances of problems related to time constrains. None of the examples present time constrains failures since the sample time is relatively high ($10\ ms$).

Additionally, the results show that direct form realizations are not a good solution for high-order digital controllers. The controllers C, D, E, and F always present overflows, although the fixed-point format and the representable range are changed. It indicates that these high-order systems should be implemented in other structures (e.g., parallel and cascade forms, where the probability of occurrence of overflows and round-off errors may be decreased).

## V. RELATED WORK

Previous work about validation methods for control systems related to FWL implementation are mostly based on simulations methods. Chattopadhyay describes a case study about the occurrence of limit cycles at DC-DC converters that employs digital current mode control and pulse-width modulation (PWM) [25]. Here, the author proposes a solution for the oscillations by adjusting the ADC resolution and the limit cycle corrector. Chattopadhyay uses the MATLAB/SIMULINK tool to verify the limit cycle and then validate the implementation. However, tests carried out with pre-specified reference current do not take into account the reminiscent oscillations for the various different current values.

Qu and Yourui propose an interesting method for PID controllers' implementations in FPGAs, with fixed-point [27]. In this work, the design of control system is carried out in SIMULINK and simulated in Modelsim [26]. The plant's behavior, after applying the method, presents stability and expected responses. However, the authors do not present any evaluation in terms of performance and error detection, which makes it difficult to compare their approach with others.

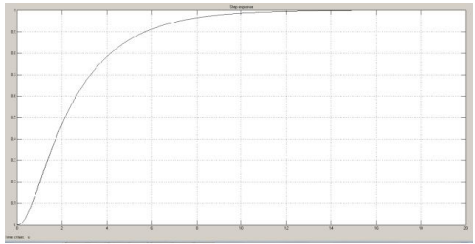

Fig. 4. Step response without quantiztions effects

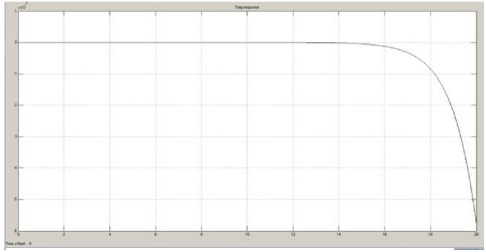

Fig. 5. Step response with quantiztions effects

Mohta [17] demonstrates that traditional design tools (e.g., SIMULINK) cannot help enough in FWL related problems and suggests the creation of a tool to determine the best FWL format implementation (i.e., coefficient word-length) to make the design process easier. Mohta presents a tool to optimize the word-length in FWL implementations, where the search for the optimal implementation uses brute force in a simulation-based environment. However, simulation approaches cannot cover all possible scenarios, as previously described. Sung and Kum also present a tool that verifies control-systems with fixed-point implementations, by searching the consequences in system stability [6]; however, it is a plant model-based tool, and models present parametric uncertainties.

Differently from others, the proposed method verifies the controllers' model without parameters uncertainties; and the use of a BMC tool ensures the absence of overflows, limit cycles, and stability problems in addition to help define the word-length of the digital controllers' implementation.

## VI. CONCLUSIONS

This paper describes a novel method to verify digital controllers, where an SMT-based BMC tool is used to verify fixed-points realizations properties of digital controllers and to identify failures that are hard to be detected by simulation tools. Digital controllers for a ball and beam plant are used to verify the occurrence of typical problems of finite-word implementation; in particular, overflow, underflow, and limit cycles. Furthermore, stability and time constraints are verified using different types of controllers' realization. The proposed method can be used as an interactive process, where controllers are firstly designed in a mathematical tool and translated into a C model; then check whether properties hold in the controller's model using a BMC tool, and repeat this process until the controller is immune to overflows and limit cycles occurrences to ensure the system's stability. The experimental results show that the stability and time constraints checks are relatively fast, while overflow and limit cycle tend to take much longer for high-order digital controllers. Additionally, the proposed method can be effective to find errors and to determine the better fixed-point structure realization in digital controllers.